\documentclass[a4paper]{article}

\usepackage{INTERSPEECH2019}
\usepackage{color}
\usepackage{subcaption}
\usepackage{amsmath,graphicx,cite}
\newcommand{\TS}[1]{{}}
\newcommand{\EW}[1]{}
\def\vx{\mathbf{x}}

\title{Large-Scale Multilingual Speech Recognition  \\
       with a Streaming End-to-End Model}
\name{Anjuli Kannan*, Arindrima Datta*, Tara N. Sainath, Eugene Weinstein,
      Bhuvana Ramabhadran, Yonghui Wu, Ankur Bapna, Zhifeng Chen, Seungji Lee
      \thanks{*equal contribution}}
\address{Google, Inc.}
\email{\{anjuli,arindrimadatta,tsainath,weinstein,bhuv\}@google.com}

\begin{document}

\maketitle
\begin{abstract}

Multilingual end-to-end (E2E) models have shown great promise in expansion of
automatic speech recognition (ASR) coverage of the world's languages.
They have shown improvement over monolingual systems,
and have simplified training and serving by eliminating language-specific
acoustic, pronunciation, and language models.  This work presents an
E2E multilingual system which is equipped to operate in low-latency
interactive applications,
as well as handle a key challenge of real world data: the imbalance
in training data across languages. Using nine Indic languages, we compare a variety of
techniques, and find that a combination
of conditioning on a language vector and training language-specific adapter
layers produces the best model. The resulting E2E multilingual model achieves a
lower word error rate (WER) than both monolingual E2E models (eight of nine
languages) and monolingual conventional systems (all nine languages).

\end{abstract}

\section{Introduction}
\label{sec:intro}


Automatic speech recognition (ASR) systems that can transcribe speech in
multiple languages, known as multilingual models, have gained popularity as
an effective way to expand ASR coverage of the world's languages.
Through shared learning of model elements across languages, they have been
shown to outperform monolingual systems, particularly for those languages with
less data.
Moreover, they significantly simplify infrastructure by supporting $n$ languages
with a single speech model rather than $n$ individual models.
Successful strategies for building multilingual acoustic models (AMs) include stacked
bottleneck features ~\cite{thomas, tuske, cui2015, sercu2017},
shared hidden layers ~\cite{ghoshal, heigold}, knowledge distillation
\cite{cui2017}, and multitask learning \cite{chen}.
Building multilingual language models (LMs) has also been attempted recently (e.g.
\cite{fugen2003efficient}).
However, in most state-of-the-art multilingual systems,
only the AM is multilingual; separate language-specific
LMs (and often lexicons) are still required.


More recently, end-to-end (E2E) multilingual systems have gained traction as a
way to further simplify the training and serving of such models.  These models
replace the acoustic, pronunciation, and language models of $n$ different
languages with a \emph{single} model while continuing to show improved
performance over monolingual E2E systems ~\cite{watanabe, toshniwal,
karafiat, cho}. Even as these E2E systems have shown promising
results, it has not been conclusively demonstrated that they can be
competitive with state-of-the-art conventional
models, nor that they can do so while still operating within the real-time constraints
of interactive applications such as a speech-enabled assistant.


Our work is motivated by the need for E2E multilingual systems that
(1) meet the latency constraints of interactive applications,
(2) handle the challenges inherent in large-scale, real-world data, and
(3) are competitive with state-of-the-art conventional models while still
operating within the same training and serving constraints.

First, we present a streaming E2E multilingual system using the Recurrent
Neural Network Transducer (RNN-T) \cite{graves}.  As \cite{Ryan19} has shown,
the architecture we employ adheres to the latency constraints required for
interactive applications.  In constrast, prior E2E multilingual work
has been limited to attention-based models that do not admit a
straightforward streaming implementation \cite{watanabe, toshniwal,
karafiat, cho}.

Next, we address the challenges of training such a model
with large-scale real world data.
Given the dramatic skew in the distribution of speakers across the world's
languages, it is typical to have varying amounts of transcribed data available
for different languages.  As a result, a multilingual model will be more
influenced by languages which are over-represented in the training set.
Working with a corpus of nine Indian languages comprised of ~37K hours of data,
we compare several techniques to address this issue: conditioning
on a language vector, adjusting language sampling ratios, and language-specific
adapter modules. Our experiments demonstrate that the combination of a
language vector and adapter modules yields the best multilingual E2E system.
While previous works have investigated various aspects of data sampling
\cite{alumae, sercu2016}, as well as architectures
that include a language vector \cite{toshniwal, li, grace}, this is the first
study to apply adapter modules \cite{rebuffi} to speech recognition.


Lastly, when we combine the above elements,
we show that the resulting system surpasses not only monolingual E2E
models, but also monolingual conventional systems built with
state-of-the-art AMs, lexica, and LMs. The E2E system consistently achieves
at least a 10\% relative reduction in WER on each of the nine languages,
when compared with the monolingual conventional systems.
This is demonstrated while still using comparable training and serving resources.

\section{Streaming E2E multilingual model}
\label{sec:rnnt}

A key requirement for an interactive application is to support streaming
ASR. Thus our experiments are conducted on RNN-T \cite{graves, rao}, a
streaming E2E
model which has been shown to offer appropriate user latency required for
applications such as a speech-enabled assistant \cite{Ryan19}.

RNN-T consists of an encoder network,
a prediction network, and a joint network.
The encoder, which is analogous to the AM in a traditional ASR system, is
a recurrent network
composed of stacked long short-term memory (LSTM) layers.
It reads a sequence of $d$-dimensional feature vectors $\mathbf{x} =
(\mathbf{x}_1, \mathbf{x}_2, \cdots, \mathbf{x}_T)$, where $\mathbf{x}_t \in
\mathbb{R}^d$, and produces at each timestep a higher-order feature
representation, denoted
${\mathbf{h}_1^{\text{enc}}}, \cdots, {\mathbf{h}_T^{\text{enc}}}$.
Similarly, the prediction network is also an LSTM network, which, like an
LM, processes the sequence of non-blank symbols output so far,
$y_0, \ldots, y_{u_{i-1}}$ into a dense
representation ${\mathbf{h}_{u_{i}}^{\text{dec}}}$.

Finally the representations produced by the encoder and prediction networks
are combined by the joint network.  The joint
network then predicts
$P(y_{i} | \vx_1, \cdots, \vx_{t_i}, y_0, \ldots, y_{u_{i-1}})$,
a distribution over the next output symbol.

In this way, RNN-T does not make a conditional independence assumption:
the prediction of each symbol is conditioned not only on the acoustics
but also on the sequence of labels output so far.  However, RNN-T does assume
an output symbol is independent of future acoustic frames.  This assumption
allows us to employ RNN-T in a streaming fashion.

\section{Imbalanced multilingual data}

This section describes three strategies we investigate for handling
data imbalance in the multilingual model. Imbalance is a natural
consequence of the varied distribution of speakers across the world's
languages. Languages with more speakers tend to produce transcribed data
more easily. In conventional ASR systems, only the AM
is trained on the transcribed speech data, but in an E2E multilingual model
all the components are trained on it.  As a result, the latter may be
more sensitive to  data imbalance.
In this section we explore two avenues to address data imbalance: (1) data
sampling and (2) extensions to the model architecture.

\subsection{Data sampling}

Imbalanced data typically leads to having a model perform better on languages
with larger data. For instance, suppose a multilingual
model is trained on $k$ languages $L_0, \cdots, L_{k-1}$,
where $L_i$ has $n_i$ training examples, and
$N = \sum_{i = 0}^{k} n_i$. At each step of training, we
assemble a batch of examples by sampling from the $N$ total examples.
Assuming the training examples have been pooled across languages and shuffled,
we expect the sampling ratio of language $L_i$ within the
batch to be $s(i) = \frac{n_i}{N}$.
This means that the model gets updated with
$\frac{n_i}{n_j}$ times more gradients generated by language $L_i$ than
by language $L_j$ for any $i, j$ where $n_i > n_j$.

One approach to dealing with data imbalance is to upsample data from
under-represented languages so that the
distribution across different languages is more even.
In the extreme case, we can
sample each language uniformly, such that $s(i) = \frac{1}{k}$ for all $i$.
More generally, we can let

\begin{equation}
s(i) = \frac{n_i + \alpha * (n^*-n_i)}{\sum_{i} [n_i + \alpha * (n^*-n_i)]}
\end{equation}

\noindent where $n^*$ is the maximum number of examples for any language and $\alpha$ is
a tunable parameter. If $\alpha = 0$ we are sampling at natural frequencies,
while if $\alpha = 1$ we are sampling uniformly.

Similar strategies are described in \cite{sercu2016, garcia}, but are applied
to the sampling of phonemes rather than languages.  Moreover,
they are focused on the AM of a conventional model, whereas we investigate
an E2E setting.  \cite{alumae} has also considered scalars within the loss
function to increase the contribution of under-represented languages, when
learning multilingual bottleneck features.

\subsection{Conditioning on language vector}

Next we explore the
architectural extension of feeding a vector representing the language.
The intuition is that a universal model can use a language vector to
become an ``expert'' on each individual language instead of learning a
representation tailored to languages with more data.  At inference time, we
assume the language is either specfied in the user's preferences, or
determined automatically from a language identification system.

Various methods of using a language vector have been previously
described and directly compared in non-streaming E2E multilingual
\cite{toshniwal} and multidialect \cite{li} models.
The language itself can be represented in
several different ways (as a one-hot vector, as an embedding vector,
or as a combination of clusters learned through cluster adaptive training (CAT)
\cite{tan}), but prior work \cite{li, grace} has
shown that the simple approach of a one-hot vector performs as well as and
sometimes better than the more complex methods.
Additionally, given such a language vector, there are many different points
in the model where it can be inserted, but our
experiments have confirmed earlier observations \cite{li, toshniwal}
that simply concatenating it to the input features is sufficient.
Thus, for all language vector experiments we represent
the language as a one-hot vector, and concatenate it to the input features
of the encoder network.

While previous works have investigated the use of the language vector in
multilingual modeling,
these have done so in the context of either the Listen, Attend and Spell
\cite{Chan15} architecture \cite{toshniwal, li}
which is a non-streaming E2E model, or conventional ASR systems
\cite{grace}.
Here we explore using a language vector with a streaming RNN-T model
and as well as its ability to deal with data imbalance.

\subsection{Adapter modules}

\begin{figure}
\centering
  \begin{subfigure}{0.25\textwidth}
  \includegraphics[width=\linewidth]{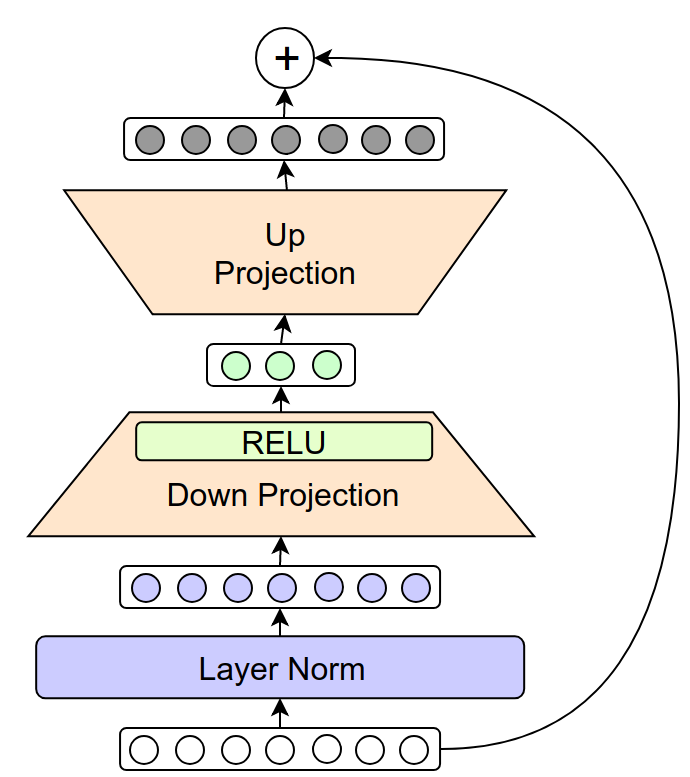}
  \caption{}
  \label{fig:adapter_module}
  \end{subfigure}
  \hspace*{\fill}
  \begin{subfigure}{0.19\textwidth}
  \includegraphics[width=\linewidth]{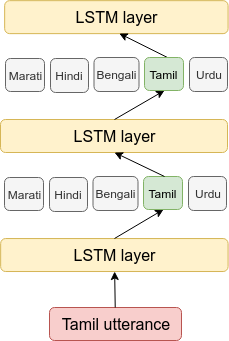}
  \caption{}
  \label{fig:adapter_encoder}
  \end{subfigure}
\caption{(a) An adapter module consists of layer normalization,
         down-projection,
         non-linearity, and up-projection. (b) Adapter modules in the context
         of the encoder.  For a Tamil utterance, only the Tamil adapters
         are applied to each activation.}
\label{fig:adapters}
\vspace{-0.15 in}
\end{figure}

A second architecture extension that we investigate to handle data imbalance is
adapter modules.
This modeling technique has seen success for domain
and task adaptation in computer
vision \cite{rebuffi}, natural language processing \cite{houlsby} and machine
translation \cite{bapna}. Here we extend it to multilingual speech recognition.

Adapter modules are effectively domain-specific (language-specific in our case)
adjustments to the activations coming out of each layer.  They aim to capture
the quality benefits of fine-tuning a global model on each language while
maintaining the parameter efficiency of having a single global model.
This is accomplished by a two-stage training process.  In the first stage, a
global model (the RNN-T described in Section \ref{sec:rnnt})
is trained on the union of data from all languages.
In a second adaptation stage, we freeze all model parameters,
and introduce  adapter modules after every
layer of the encoder. Importantly, each adapter module contains
separate parameters for each language.

More details are shown in Figure \ref{fig:adapters}.  Figure
\ref{fig:adapter_module} shows the exact computation of each module:
following \cite{houlsby, bapna}, each activation is projected down to a
smaller dimensionality, passed through a non-linearity,
then projected back up to the original size.  This result is then added to
the original activation before being passed to the next layer.
Figure \ref{fig:adapter_encoder} shows what happens at inference time: a
Tamil utterance comes in, and thus the Tamil-specific adapter weights are
applied after each layer.

The result of this training can be viewed as a single multilingual model with
a small number of language-specific parameters (typically less than
10\% of the original model size). This allows for efficient
parameter sharing across languages, as well as per-language specialization.

A conceptually similar technique called learning hidden unit contributions
(LHUC) \cite{swietojanski} has been successfuly applied to multilingual ASR by
\cite{tong}.  The main difference is that LHUC modulates the amplitude of
the different hidden units, whereas adapter modules are entirely separate
layers.  \cite{bapna} has shown adapter modules to be more effective than
LHUC for large-scale sequence-to-sequence models, so we limit our study to this
method.

\section{Experimental details}
\label{sec:experiments}
\vspace{-0.025 in}
\subsection{Data}

\begin{table}[!t]
\caption{Number of utterances in train and test sets}
\vspace{-0.1 in}
\label{table:data}
\centering
\begin{tabular}{c|c c||c|c c}
\toprule
Language  & Train & Test   &  Language & Train & Test \\ \hline
Hindi     &  16M  & 6.3K  & Tamil     &  1.8M & 5.5K \\
Marathi   &  4.1M  & 6.1K  & Malayalam &  1.5M   & 9.2K \\
Bengali   &  3.9M   & 3.6K & Kannada   &  1.2M   & 1.1K \\
Telegu    &  2.4M  & 2.7K  & Urdu      &  443K   & 511  \\
Gujarati  &  2.2M  & 7.5K  & \textbf{Total}  &  33M  & 43K \\

\bottomrule
\end{tabular}
\vspace{-0.2 in}
\end{table}

Our training data and test data consist of anonymized,
human-transcribed utterances, representative of Google's traffic,
and spanning nine Indian languages. Train and test set sizes vary due
to availability of transcribed data and are shown in Table \ref{table:data}.
Training data are further augmented by corrupting clean utterances
using a room simulator \cite{Chanwoo17}.

Each of the nine languages is written in a different script, except Hindi
and Marathi, which both use the Devanagari writing system.  In addition, each
language's transcriptions have some Latin alphabet mixed in, as transcribers
may use Latin alphabet for loaner words and some proper nouns.  The frequency of
Latin alphabet is similar to what is reported in \cite{emond}.

Since transcripts can contain a mixture of Latin and native scripts, the metric
we report is \textit{transliteration-optimized WER} \cite{emond}.  Put
simply, this means that if the model correctly decodes a word in Latin
alphabet but the reference is in the native script (or vice versa), this
is not considered an error, because the output can be transliterated
to one or the other for rendering.

\vspace{-0.025 in}
\subsection{Model architecture}

All experiments use 80-dimensional log-mel features, computed with a 25ms
window and shifted every 10ms. These features are stacked with 7 frames to the
left and downsampled to 30ms frame rate.

The encoder network consists of eight 2,048-dimensional LSTM layers, each
followed by a 640-dimensional projection layer. The prediction network
has two 2,048-dimensional LSTM layers, each of which is also followed by
640-dimensional projection layer. Finally, the joint network also has
640 hidden units. The softmax layer is composed of a
unified grapheme set from all languages (988 graphemes in total),
that was generated using all unique graphemes in the training data.
Adapter modules use a 256-dimensional bottlneck after each of the eight
encoder layers.

All RNN-T models are trained in Lingvo \cite{shen2019lingvo} on
$8 \times 8$ Tensor Processing Units \cite{tpu} slices with a batch size of
4,096.

\vspace{-0.025 in}
\subsection{Baselines}

We report two baselines for comparison.  First, RNN-T monolingual baselines are
trained on each of the individual languages using the same architecture as the
multilingual model described earlier, with the addition of L2 regularization to
reduce overfitting. Additionally, monolingual
conventional models are trained as follows.  The AMs consist of five
768-dimensional LSTM layers and a softmax over over context dependent phone
states.  They are trained using connectionist temporal
classification (CTC)\cite{graves2006connectionist}, followed by state-level
minimum Bayes risk \cite{kingsbury2009lattice,sak2014sequence}.
The AM outputs are then used with standard FST-based beam-search decoders with
language-specific lexicons and 5-gram language models for decoding.

\section{Results}

\label{sec:results}

\begin{table*}
  \centering
  \caption{WER of the multilingual E2E Model using various techniques to address data imbalance.}
  \vspace{-0.08 in}
  \begin{tabular}{c|p{2.9cm}||c|c|c|c|c|c|c|c|c||c}
  \toprule
    Exp & Model & Hindi & Marathi & Beng. & Telugu & Gujarati & Tamil & Mala. & Kann. & Urdu & Avg \\ \hline
    A0 & Multilingual RNN-T & 18.5 & 26.2 & 43.9 & 49.3 & 55.3 & 40.1 & 69.7 & 60.8 & 70.1 & 48.2\\
    A1 & A0 + language vector  & 16.0 & 17.6 & 22.8 & 23.5 & 24.3 & 22.2 & 46.6 & 20.5 & 17.3 & 22.8 \\
    A2 & A0 + sampling  & 22.3 & 29.8 & 41.1 & 45.9 & 43.9 & 37.7 & 64.6 & 55.4 & 48.1 & 43.2 \\
    A3 & A1 + sampling @ 60K  & 18.7 & 18.8 & 24.0 & 24.6 & 24.3 & 25.0 & 47.8 & 21.4 & 17.7 & 24.7 \\
    A4 & A1 + sampling  & 16.2 & 17.8 & 24.1 & 25.1 & 24.2 & 22.9 & 48.9 & 24.6 & 20.4 & 24.9 \\
    A5 & A1 + adapters  & \textbf{15.9} & \textbf{17.1} & \textbf{21.5} & \textbf{23.2} & \textbf{24.0} & \textbf{21.6} & \textbf{45.8} & \textbf{18.7} & \textbf{16.0} & \textbf{22.6} \\
  \bottomrule
  \end{tabular}
  \label{table:comparison}
\vspace{-0.05 in}
\end{table*}

\begin{table*}
\vspace{-0.05 in}
\centering
  \caption{WER from the best multilingual E2E model and monolingual baselines.}
  \vspace{-0.08 in}
  \begin{tabular}{c|p{2.9cm}||c|c|c|c|c|c|c|c|c||c}
  \toprule
    Exp & Model & Hindi & Marathi & Beng. & Telugu & Gujarati & Tamil & Mala. & Kann. & Urdu & Avg \\ \hline
    B0 & Monolingual CTC  & 18.6 & 19.8 & 26.8 & 25.1 & 29.6 & 24.5 & 47.1 & 30.0 & 29.5 & 28.0 \\
    B1 & Monolingual RNN-T & 16.1 & 21.3 & \textbf{18.2} & 25.5 & 26.4 & 27.6 & 54.4 & 29.5 & 27.4 & 27.4 \\
    B2 & Multilingual RNN-T & \textbf{15.9} & \textbf{17.1} & 21.5 & \textbf{23.2} & \textbf{24.0} & \textbf{21.6} & \textbf{45.8} & \textbf{18.7} & \textbf{16.0} & \textbf{22.6} \\
  \bottomrule
  \end{tabular}
  \label{table:comparison}
  \vspace{0.05 in}
  \emph{Languages are listed in descending order of training data amount.}
  \vspace{-0.1 in}
\end{table*}

\subsection{Conditioning on language vector}
\label{sec:langid}

We begin by investigating the impact of conditioning the RNN-T encoder on a
language vector.  As shown in Table \ref{table:comparison} (comparing
\texttt{A0} and \texttt{A1}), providing this information to the model is
critical to good performance,
particularly for languages with less data such as Urdu, Kannada, and Malayalam,
which all see more than 50\% relative WER reduction.

This result is somewhat surprising, given previous works have shown a
much smaller impact of providing language information \cite{li, toshniwal}.
However, our error analysis reveals that this is largely a result of the
significant overlap in vocabulary between different languages in our data.
Because our data comes from South Asia, we see that the different languages'
corpora contain many of the same proper nouns and English loaner words.
Moreover, most utterances in our test set are short and
contain few glue words which could help with language disambiguation.
Queries such as ``arundhati movie'', ``train bangalore'', and
``doctor rajkumar'' could appear in any of the languages' training or test sets.

As a result, the model \texttt{A0} often defaults to the script
of the dominant language in the training data, Hindi.
For instance, about 75\% utterances in the Bengali
test set consist entirely of proper nouns or English words,
and 44\% of these are decoded by model \texttt{A0} in an incorrect script
but are otherwise correct.
On the other hand, model \texttt{A1} can use the language vector to ensure the
correct script is output: its Bengali WER is nearly 50\% lower than that of
\texttt{A0}.

Given the extent to which these languages' vocabularies overlap,
it is thus of little surprise that the language
vector is such a critical piece of information for this model.  What is more
notable is how this vector also helps to address the issue of data imbalance,
as we demonstrate in the the following section.

\subsection{Data sampling}
\label{sec:sampling_results}

Next we compare models with various sampling strategies.  First, we increase
$\alpha$ from 0 to 0.25 on model \texttt{A0}, which means we are upsampling
smaller languages. Expectedly, we see that the WER on all but the two largest
languages (Hindi and Marathi) decreases (compare \texttt{A2} to \texttt{A0}).
We effectively
change the model's prior on what language to output, so that when it
comes across an ambiguous utterance, it may be less likely to default
to the dominant language.  However, this does result in a 10-20\% relative
regression on the two largest languages.

When we repeat this comparison on the model with the language vector
(\texttt{A1}), we find that all benefit of upsampling
has been eliminated (compare \texttt{A4} and \texttt{A1}).
While the upsampled languages do reach their lowest WER
earlier in training, it never drops below their WER in \texttt{A1}.
Moreover, the upsampled languages begin to overfit while larger
languages keep improving.

This phenomenon is demonstrated by comparing \texttt{A3} and \texttt{A4}
against \texttt{A1}.  \texttt{A3}
shows this model evaluated mid-way through training (approximately 60K steps)
when small languages like
Kannada and Urdu have hit their lowest WER.  At this stage, we see that large
languages (Hindi, Marathi) are still underfitting the data compared with
\texttt{A1}. Yet by the time they have fully fit the data (\texttt{A4}) the
small languages have heavily overfit.  This phenomenon is only exacerbated
by increasing $\alpha$.

We hypothesize that the model is able to use the language vector not only
to disambiguate the language (as we discussed in the previous section) but
also to learn separate features for separate languages, as needed.  Thus,
upsampling a small language like Kannada does not help the model to
learn a better representation of Kannada speech, as the model is already
allotting the necessary capacity.

\vspace{-0.04 in}
\subsection{Adapters}

Lastly, we investigate adapter modules as an architectural
modification to address imbalance between languages. Comparing \texttt{A5}
to \texttt{A1} we see that the adapters
provide a small additional reduction in WER on all languages,
with the largest gains on  Kannada (9\% relative), Urdu (8\% relative), and
Bengali (6\% relative).
We hypothesize that the encoder of the global
model is mostly shaped by the dominant languages, Hindi and Marathi,
whereas Kannada, Urdu, and Bengali may have distinct acoustic features that can
be captured by small per-layer adjustments.

Adding language-specific adapter modules allows the model to specialize in each
language in the same way that fine-tuning the whole model would, but in a
much more parameter-efficient way: the capacity added for each
language (2.5M parameters) is only about two percent of the original model size
(120M parameters).  We also point out that adapters do not need to be employed
when they are not helpful.
For example, in practice we might choose to only keep adapters for Kannada,
Urdu, and Bengali, where they are most effective, while setting all other
adapters to identity operations (i.e., setting the weights to zeroes).
This model would only be 6\% larger than the original model while
showing significant gains on those three languages.

\vspace{-0.07 in}
\subsection{Comparison with Baseline Models}

Combining the above results, we compare our best multilingual RNN-T model
against both monolingual RNN-T baselines and monolingual conventional baselines.
We observe that the multilingual model has a lower WER than the
monolingual RNN-T on eight of the nine languages.
Similarly, it achieves a lower WER than the monolingual conventional models on
all nine languages, consistently by about 10\% relative, with larger relative gains
(34\% on Kannada and 25\% on Urdu) on the lower resource languages.  Future
work will be needed to understand the role that shared vocabulary and phonetic
relatedness plays in this result.

In addition to a lower WER, the multilingual RNN-T model offers the
benefit of replacing nine separate recognizers (each of which typically consists
of acoustic, pronunciation, and language models) with a single, compact
recognizer.
To our knowledge, this is the first work
to demonstrate that a multilingual E2E model which is suitable for streaming
applications can outperform monolingual conventional models.

\vspace{-0.04 in}

\section{Conclusions}
\label{sec:conclusion}
\vspace{-0.04 in}

In this work, we extended prior work on E2E multilingual models to address
two issues that arise in large-scale practical applications:
(1) the need for streaming ASR and
(2) the challenge of imbalanced training data.
We presented a system that addresses both issues, as well as a comparison of
techniques to address the second.
Using nine Indian languages, we showed that our best system, built with
RNN-T model and adapter modules, significantly outperforms both the monolingual RNN-T models, and
the state-of-the-art monolingual conventional recognizers.

\vspace{-0.07 in}
\section{Acknowledgements}
\label{sec:acknowledgements}

The authors would like to thank Ruoming Pang, Sergey Kishchenko,
Pedro Moreno, Mikaela Grace, Bo Li, and Meysam Bastani for helpful discussions.

\bibliographystyle{IEEEtran}

\bibliography{paper}

\end{document}